\documentclass{emulateapj}

\submitted{Submitted on 2006 April 14}

\usepackage{amsmath}
\usepackage{xspace} 
\usepackage{graphicx}
\usepackage{txfonts}
\usepackage{natbib}

\long\def\symbolfootnote[#1]#2{\begingroup\def\thefootnote{\fnsymbol{footnote}}\footnote[#1]{#2}\endgroup}

\newcommand{\un}[1]{\,\ensuremath{\mathrm{#1}}}
\newcommand{\as}{$^{\prime\prime}$\xspace}

\newcommand{\msun}{\,$M_{\odot}$\xspace}
\newcommand{\msol}{\msun}

\newcommand{\chandra}{{\em Chandra}\xspace}
\newcommand{\xmm}{{\em XMM-Newton}\xspace}

\newcommand{\mos}{MOS\xspace}

\newcommand{\pn}{PN\xspace}

\newcommand{\sgra}{Sgr\,A$^{*}$\xspace}

\newcommand{\xray}{X-ray\xspace}
\newcommand{\xrays}{X-rays\xspace}

\newcommand{\nir}{near-IR\xspace}

\newcommand{\mso}{MSO\xspace}

\newcommand{\pdf}{PDF\xspace}

\begin{document}

\title{Periodic Modulations in an X-ray Flare from Sagittarius A*}

\author{G.\ B\'elanger\altaffilmark{1,2}, R.\ Terrier\altaffilmark{1,2}, 
	O.C.\ de Jager\altaffilmark{3},
	A.\ Goldwurm\altaffilmark{1,2},
	F.\ Melia\altaffilmark{4}
	}

\altaffiltext{1}{\scriptsize SAp, DAPNIA/DSM/CEA, 
			91191 Gif-sur-Yvette, France; belanger@cea.fr}
\altaffiltext{2}{\scriptsize UMR Astroparticule et Cosmologie, 
			11 place Berthelot, 75005 Paris, France}
\altaffiltext{3}{\scriptsize Unit for Space Physics, North-West University,
			Potchefstroom 2520, South Africa}
\altaffiltext{4}{\scriptsize Physics Dept. and Stewart Observatory, University of Arizona, 
			Tucson, AZ 85721, USA}

\begin{abstract}
We present the highly significant detection of  a quasi-periodic flux modulation 
with a period of 22.2\un{min} seen in the \xray data of the \sgra flare of 2004 August 31. 
This flaring event, which lasted a total of about three hours, was detected 
simultaneously by EPIC on \xmm and the NICMOS near-infrared camera on the HST.
Given the inherent difficulty in, and the lack of readily available methods for 
quantifying the probability of a periodic signal detected over only several cycles 
in a data set where red noise can be important, we developed a general method 
for quantifying the likelihood that such a modulation is indeed intrinsic to the source
and does not arise from background fluctuations. We here describe this Monte Carlo based 
method, and discuss the results obtained by its application to a other \xmm data sets.
Under the simplest hypothesis that we witnessed a transient event that evolved,
peaked and decayed near the marginally stable orbit of the supermassive black hole, 
this result implies that for a mass of 3.5\,$\times$\,$10^{6}$\msol, the central object 
must have an angular momentum corresponding to a spin parameter of $a$\,$\approx$\,0.22.  
\end{abstract}

\keywords{black hole physics --- Galaxy: center --- Galaxy: nucleus --- accretion --- 
	\xrays: observations --- Methods: data analysis}

\section{Introduction}
\label{s:intro}


Detecting Keplerian motion in the accretion flow orbiting around a black hole (BH),
in the case where an event, localized to a portion of the flow, retains its coherence
over several orbits, entails being able to detect periods of the order of $10^{-4}$\un{s}
(or frequencies of $\sim$10\un{kHz}) for solar mass objects. The problem is entirely 
different when considering a supermassive black hole (SMBH), and in particular \sgra.
For a non-rotating BH of 3.5\,$\times$\,$10^{6}$\msol, the period at the 
marginally stable orbit (\mso, $r$({\sc mso\/})\,=\,3$r_{\rm s}$\,=\,$6GM/c^2$) 
is 1592\un{s}  or about 26.5\un{min}. So as much as 
detecting kHz quasi-periodic oscillations may be a delicate matter for certain reasons,
as we explore long periodic signals, low frequency noise and windowing
effects become important, and must be taken into account in a proper 
characterization of the signal.

In general, to test for a periodic signal, we must know what is the probability 
distribution function (\pdf) of the test statistic in the absence of such a periodic signal,
so that a reliable significance can be assigned to the claimed detection.
In Fourier analysis, a \pdf of the form $e^{-Z}$, where
$Z$ is the Fourier power gives the classical probability measure.
For time series with important fluctuations or flares (colored noise), 
one usually obtains a distribution that is broader than $e^{-Z}$ resulting in 
overestimated significances.
Similarly, windowing effects introduced by the testing of non-integral periods with
respect to the total observation time also affect the probability distribution.
These effects can be taken into account in order to derive an accurate probability
for a given peak in the power spectrum.

For \sgra, the \mso---where we expect the minimum fundamental Keplerian 
period---ranges from $P_{\rm min}$\,$\sim$\,300 to $P_{\rm max}$\,$\sim$\,1600\un{s}, 
for a mass of 3.5\,$\times$\,$10^{6}$\msol
and spin parameter 0\,$\ge$\,$a$\,=\,$\frac{Jc}{GM^2}$\,$<$\,0.999. Here, $J$ is the 
angular momentum, $c$ is the speed of light, and $G$ the gravitational constant. 
A number of \xray observations of \sgra have revealed flares with durations between 
$\approx$\;3\,000 and 10\,000\un{s}.  We adopt a standard minimum quality factor, 
$Q$, of 4--5 cycles to claim a periodicity, and thus find that for the longest period of 
1600\un{s}, the minimum flare duration must be 8000\un{s}. The number of 
Independent Fourier Spacings (IFS or trials) in our full interval (8752\un{s}) is given by:
IFS\,=\,$T_{\rm obs}(1/P_{\rm min} - 1/P_{\rm max})$\,=\,24.

Ascenbach et al. (2004) claimed the detection of 5 periods ranging from 
100 to 2250\un{s} each of which were identified with one of the gravitational cyclic
modes associated with accretion disks. The work of these authors was based on 
\xray data from \xmm. Genzel et al.\ (2003) claimed the detection of an
$\approx$\,17\un{min} period detected during two \nir flares from \sgra.
Clearly, a confirmed detection of a periodic or quasi-period signal present in a flare
from the central BH would be very useful for constraining the nature of the flares,
and more importantly, as evidence for the geometry of the space-time around \sgra
through indirect BH spin measurements.
In the context of this paper, we use the term {\em quasi-period} to refer to the
average period of a signal that exhibits a periodic quality, while possibly 
comprising an evolution from longer to shorter periods, which would
leads to a broader peak due to a spread in the power over several frequencies.

\section{Observations and Methods}
\label{s:methods}

\subsection{Observations}

In the last few years, observations of the central BH with the \chandra 
and \xmm \xray telescopes have revealed several flares with different durations 
and spectral indices \citep{c:baganoff01, c:baganoff03, c:goldwurm03, c:porquet03}. 
Recently, two more flares were detected by \xmm and their properties are 
discussed by B\'elanger et al.\ (2005).

Here, we restrict our analysis to \xmm data, and pay particular attention to two data sets:
0516-0111350301 and 0866-0202670701.
These observations of \sgra were respectively performed on 2002 October 3
and 2004 August 31, and each comprises one major flare, that lasted
$\sim$3 and 9\un{ks} respectively.
Details on the observation and characteristics of the first flare are presented 
and discussed by Porquet et al.\ (2003), and those of the other event by 
B\'elanger et al.\ (2005). In addition,  we analysed the data of pointing 
0789-0202670601, which also comprised a flare from \sgra, but solar proton 
contamination caused a large number of data gaps during the flaring events 
thus making these data unreliable for period analysis. As described in 
B\'elanger et al.\ (2005), the light curve can be reconstructed under the 
condition that the binning is large enough to be insensitive to the presence 
of very short good time intervals that would otherwise lead to an overestimate 
of the count rate. In this case, 500\un{s} was found to be a reasonable bin size, 
but this does not allow a meaningful period analysis for an event that lasted
$\sim$3\un{ks}.

\begin{figure}[htb]
\epsscale{1.0}
\begin{center}
	\includegraphics[scale=0.28, angle=-90]{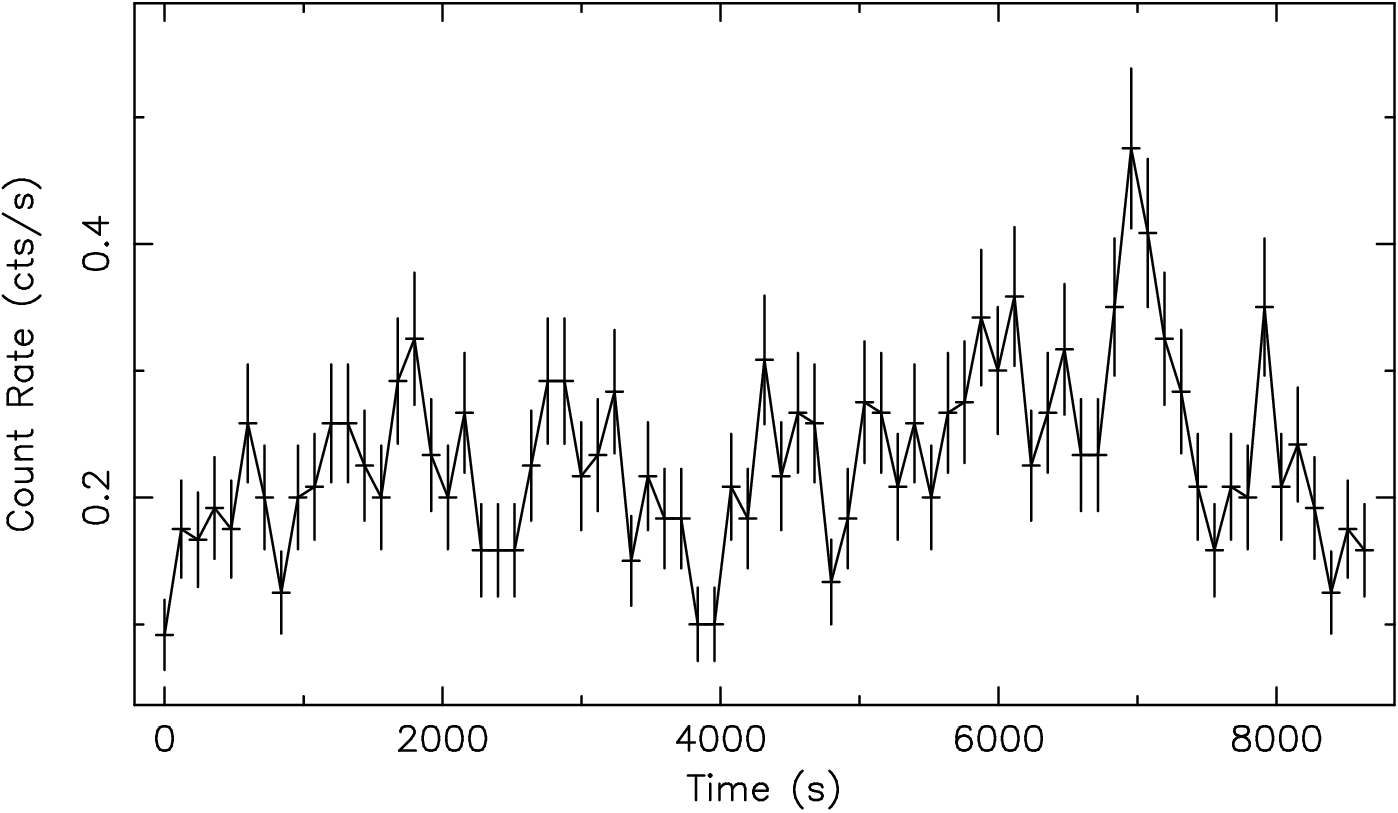}

	\vspace{3mm}

	\includegraphics[scale=0.28, angle=-90]{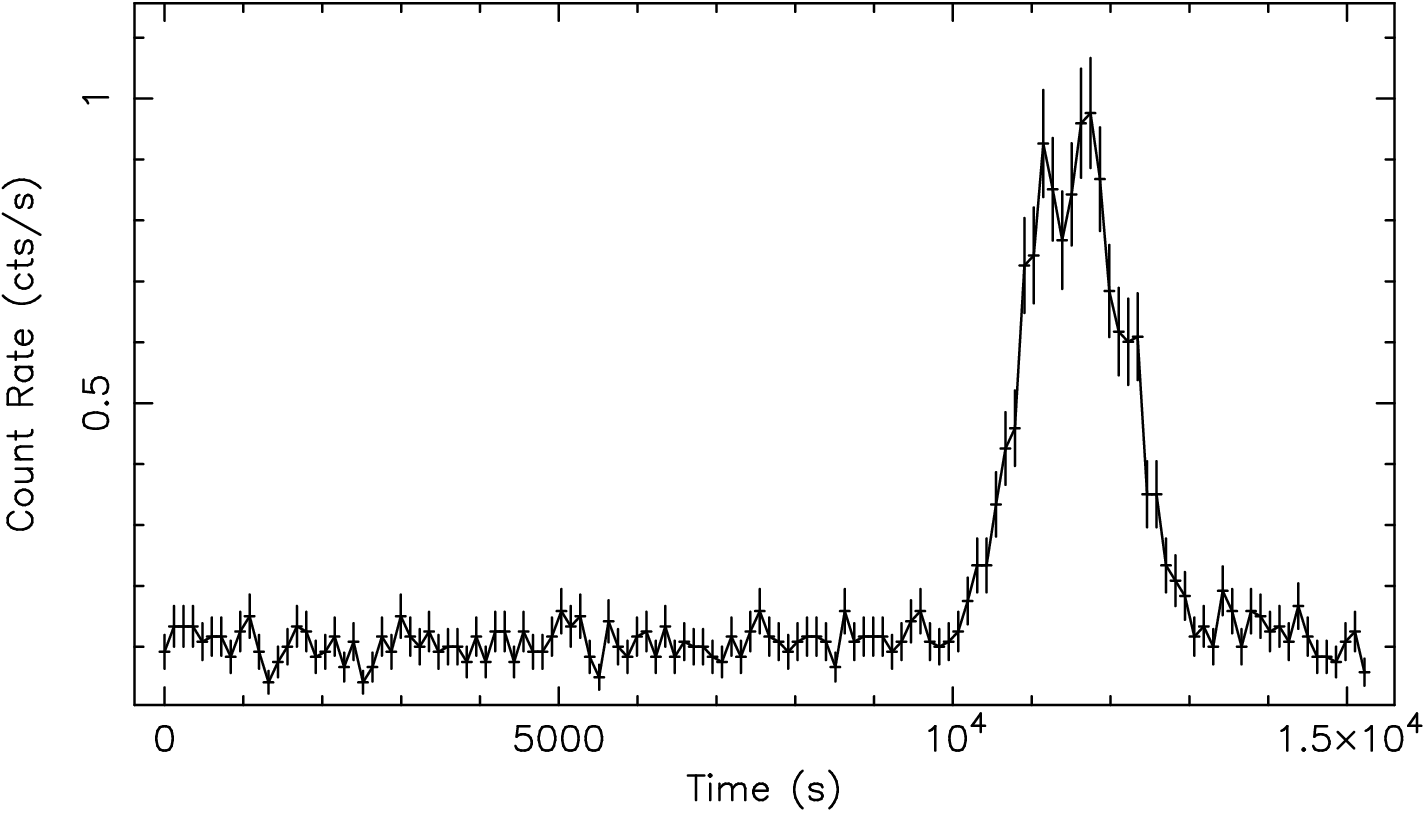}
\end{center}\vspace{-3mm}
	\caption{\scriptsize {\em Top panel:} Light curve of the 2004 August 31
	flare ($T$\,$\sim$\,9\un{ks}). {\em Bottom panel:} Light
	curve of complete $\sim$15\un{ks} 2002 October 3 observation.
	Resolution is 120\un{s}.}
        \label{f:lightcurves}
\end{figure}

For each data set, we used the \pn event list constructed by selecting good 
\xray events in the 2--10\un{keV} energy range, within a radial distance of 10\as 
from the radio position of \sgra \citep{c:yusefZadeh99}. The \pn instrument has a 
larger collecting surface, better sensitivity and most importantly, a much higher
time resolution than the \mos cameras---73\un{ms} compared to 2.6\un{s} in 
{\em FullWindow} mode%
\footnote{\scriptsize%
	\pn's time resolution can be as high as 0.3\un{ms} in the 
	standard {\em Timing} mode with a duty cycle close to 100\%, and 
	7\,$\mu$s in the special {\em Burst} mode. The duty cycle, however, 
	is only 3\% in the latter mode of operation.
}.
This makes the \pn camera the \xmm instrument of choice for timing analysis.
Figure\;\ref{f:lightcurves} show the light curve of the 2004 August 31 
flare and of the 2002 October 3 observations, both with 120\un{s} resolution.

\subsection{Test-statistic and Periodogram}
\label{s:thePeriodogram}

The $Z^2$-periodogram is constructed as follows.
For a given test period $P_j$, the phase $\phi_i$, of each arrival time $t_i$,
is calculated as $\phi_i$\,=\,$2\pi t_i/P_j$.
These phases are then used to calculate the $Z^2_m$ value
defined by Buccheri et al.\ 1983 as
\begin{equation}
	Z^2_m = \frac{2}{N} \sum_{k=1}^{m}
	\left[
		\left(\sum_{i=1}^{N}\cos(k\phi_i)\right)^2 
		+ \left(\sum_{i=1}^{N}\sin(k\phi_i)\right)^2
	\right] 
\end{equation}
Here, $N$ is the number of events and $m$ is the number of harmonics over 
which the sum is performed. We used the Rayleigh statistic ($m$\,=\,1) and thus performed
the sum over the first harmonic only, since this is the most powerful test for sinusoidal
signals (Leahy, Elsner \& Weisskopf 1983).

The Rayleigh statistic, $Z^2_1$ is distributed as a $\chi^2$ with two degrees of freedom (dof).
This is so because $\cos\phi$ and $\sin\phi$ are Gaussian distributed, the square of a
Gaussian variable is $chi^2_1$ distributed, and the sum of $\chi^2$ variables is also
a $chi^2$ variable for which the number of dof is given by the sum of the dof from the 
individual variables. Thus, the $Z^2_m$ statistic is $\chi^2_2m$ distributed.
de Jager (1994) showed that the predicted significance for a sinusoid from the
Rayleigh power is 
\begin{equation}
Y = - \log_{10} ({\rm Prob}) = 0.434(Np^{2}/4 + 1), 
\label{eq:deJager94}
\end{equation}
where $N$ is the number of events, and $p$ is the pulsed fraction.

As mentioned above, the range of periods related to the \mso we consider physically 
interesting is 300--1600\un{s}. Nonetheless, periodograms are shown from 100\un{s} 
to $T_{\rm obs}/3$.

\subsection{Background}
\label{s:background}

In the simplest case, the background is composed of white or Poisson noise:
fluctuations for which the range of values is consistent with counting statistics.
A more general case considers the background as a combination of white and red noise. 
Red noise (sometimes referred to as $1/f$ noise) arises from the inherent source variability
 and has a power spectral distribution that goes as $f^{-\alpha}$, where $f$ is the 
frequency and $\alpha$ is the slope of the power spectrum in log-log space---the 
larger the amplitude of the variations, the steeper the slope.

Clearly, it is essential to determine the relative importance of the component 
of red noise with respect to the Poisson noise in order to accurately estimate the 
probability associated with a given peak in the periodogram.
This can be done in different ways, one of which was recently suggested by 
Vaughan (2005) and entails determining the index $\alpha$ by performing a linear 
fit on the log-log periodogram and then calculating in an analytical way the confidence 
levels, conveniently represented as lines parallel to, and lying above the fitted power-law.
This method is simple, does not require Monte Carlo (MC) simulations, and is  
practical for data that has an important component of red noise and whose duration 
is long enough to allow for a suitably large range of frequencies with a well defined 
power spectrum over which to perform the fit.\footnote{%
	A proper fitting technique must be used in order to determine the 
	power-law index (see Goldstein, Morris \& Yen 2004).}

Another way to characterize the component of red noise in an observation is to 
construct the power spectrum of the appropriately smoothed light curve.
The power spectrum will then be dominated by the features 
associated with the inherent source variability, and will exhibit the
features of the underlying red noise component, which can then be characterized 
as described above. This procedure should give results closely
compatible with the method of de-trending---fitting the light curve with a 
polynomial and subtracting it---which leaves the flattened light curve free of the 
inherent source variability. Here, however, the stochastic nature of the red noise
is not accounted for.

We have used a method that allows us to work directly with the event list
and to take into account all noise components present in the data in a natural and 
transparent manner that also takes into account the random nature of the noise. 
This was done with the use of MC simulations to generate event lists having the
same statistical properties as the data.

\subsection{Pseudo-Random Event Lists}

Most analyses aimed at estimating the significance of a periodic signal use MC
simulations. We have used such a technique to estimate the probability of the null
hypothesis, that is, the probability that a peak in the periodogram is caused by a 
background fluctuation. However, several different methods can be employed, and 
thus we detail the one we have used.

In counting photons from a given astrophysical source or 
radioactive decays from an unstable isotope, each event occurs independently of 
the previous but with a certain regularity given by the average count rate.
These processes can be described by the same statistical law, characterised by an 
exponential \pdf that defines the distribution of time intervals between two 
consecutive events and for which the mean corresponds to the average count rate. 
Therefore, in order to construct a true pseudo-random light curve
of total duration $T$ and average count rate $r$, one must draw 
$T$\,$\times$\,$r$ numbers from an exponential \pdf with a mean of $r$.
Each of these corresponds to the $\Delta t$ between two consecutive events.
Arrival times are calculated by summing these $\Delta t$, and the light 
curve is finally constructed from an event list that has the same statistical
properties (Poissonian for $r$\,=\,const), as the data.

In the case where the mean is not constant, we can generate event lists with 
equivalent statistical properties as the data set by drawing the $\Delta t$s from 
different exponential \pdf\/s with means determined by the sliding average 
count rate calculated from the data over an appropriately sized window.

The size of the smoothing window determines the range of periods accessible,
(period longer than this size are smoothed out). By performing the 
simulations using different window sizes determined on the basis of the length 
of the observation as $T/4$, $T/8$, $T/16$, $T/32$, etc. down to a specified minimum, 
the probability associated with each trial period is determined taking into account
fluctuations in the base level of the light curve that occur on timescales of the
smoothing window size. Therefore, in each range of accessible trial periods,
the significance of each peak is properly assessed. A period will be 
picked out only if it is significant compared to general trends in the light curve.  
Drawing numbers from \pdf\/s with means that follow the variation in the count rate
takes into account the overall noise properties of the data by simultaneously
characterizing the inherent variations in the base level of the light curve, and 
allowing fluctuations about this level that are proportional to the amplitude of
the variation. This is the method we used to estimate the probability associated
with each value in the Rayleigh periodogram.

In practice, for a given window size, we determine a set of mean rates and
errors. For a mean rate $r$\,$\pm$\,$\sigma_{r}$, a $\Delta\/t$ is constructed 
in two steps: one pseudo-random number, $\hat{r}$, is drawn from a Gaussian 
N($r$,$\sigma_{r}$), and another, $\Delta\hat{t}$, is drawn from an Exponential 
E($\hat{r}$). The phases are constructed from the arrival times as described above. 
The full range of testable periods (chosen as 100--$T/3$) is subdivided in 100 
logarithmically spaced period bins, and logarithmically sampled by 150 trial 
periods, $P_j$, to ensure that each bin contains the same number of trials. 
For each $P_j$, we obtain a $Z^2$ value histogrammed in 
the corresponding period bin. 
Finally, the probability is determined directly from the $Z^2$ values
by finding the quantile inverse for the $Z^2$ value of $P_j$ in the data set.


\section{Results}
\label{s:results}

\begin{figure}[htb]
    \epsscale{1.0}
    \begin{center}
	\includegraphics[scale=0.3, angle=-90]{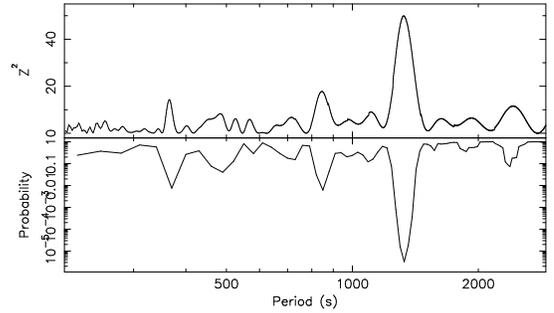}
    \end{center}\vspace{-3mm}
	\caption{\scriptsize Flare of 2004 August 31: 
	$Z^2$-periodogram and associated probabilities derived from 
	MC simulations of $10^6$ event lists. Total flare duration 
	was $\sim$9000\un{s} and thus the period range is 100--3000\un{s}.
	The peak is at $P$\,=\,1330\un{s} and rises to a $Z^2$ value of 46,
	and has an associated probability of $\sim$$10^{-6}$. }
        \label{f:flare866}
\end{figure}

\begin{figure}[htb]
    \epsscale{1.0}
    \begin{center}
	\includegraphics[scale=0.3, angle=-90]{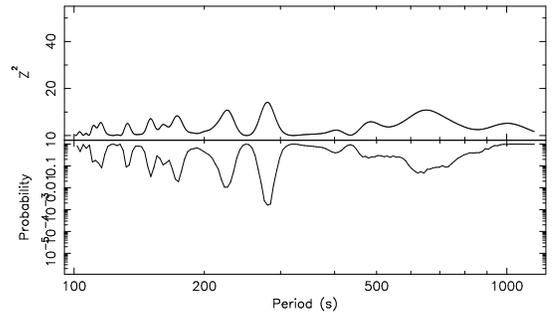}
    \end{center}\vspace{-3mm}
	\caption{\scriptsize Flare of 2002 October 3: 
	$Z^2$-periodogram and associated probabilities derived from 
	MC simulations of $500$ event lists. Total flare duration
	is $\sim$3000\un{s} and period range spans 100--1000\un{s}.
	No significant peaks are detected} 
        \label{f:flare516}
\end{figure}

\begin{figure}[htb]
    \epsscale{1.0}
    \begin{center}
	\includegraphics[scale=0.3, angle=-90]{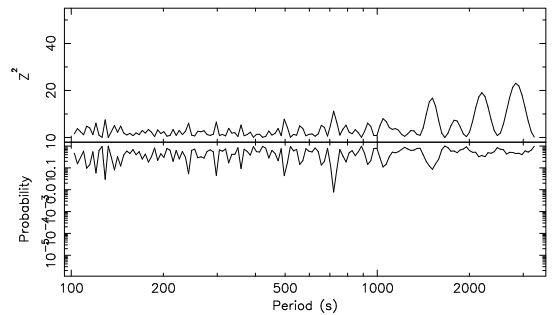}
    \end{center}\vspace{-3mm}
	\caption{\scriptsize Data of 2002 October 3 before flare: 
	$Z^2$-periodogram and associated probabilities derived from 
	MC simulations of $500$ event lists. Total duration of
	observation is $\sim$10000\un{s} and period range spans 100--3000\un{s}.
	No significant peaks are detected}
        \label{f:beforeFlare516}
\end{figure}

Figures\;\ref{f:flare866} and \ref{f:flare516} show the $Z^2$-periodograms, and their 
associated probability as a function of period derived from the MC simulations for the 
flare data sets of 2004 August 31 and and 2002 October 3 respectively. The
In Fig.\;\ref{f:beforeFlare516}, we present the periodogram and probability 
of the data obtained on 2002 October 3 before the flare. 
The vertical scale is the same on all three Figures.

As expected, the probability function is perfectly anticorrelated with the value 
of $Z^2$ over most of the range.
Deviations are apparent towards to end of the range, where $P_j$
becomes commensurable with $T_{\rm obs}$. This effect is most noticible
in Fig.\;\ref{f:beforeFlare516} where the last three peaks in the periodogram grow
in height whereas their probabilities hover around one, indicating
indicates that the MC simulations indeed model the overall statistical properties 
of the data accurately and thus reproduce features associated with the noise. 
Notice that similar behaviour, although not as marked, is also 
seen in the 2004 August 31 as well as 2003 October 3 events.

\begin{figure}[htb]
    \epsscale{1.0}
    \begin{center}
	\includegraphics[scale=0.33, angle=-90]{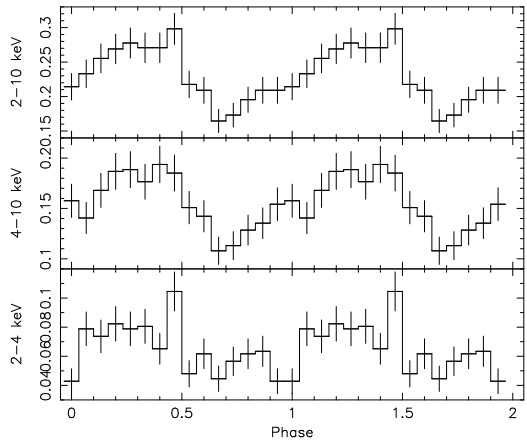}
    \end{center}\vspace{-3mm}
	\caption{\scriptsize Light curve of 2004 August 31 flare 
	folded with a phase of 1330\un{s}, in three energy bands:
	2--10\un{keV} (top), 2--4\un{keV} (middle) and 4--10\un{keV}
	(bottom).}
        \label{f:phasogram}
\end{figure}

In Fig.\;\ref{f:flare866}, the strongest peak at 1330\un{s} has an associated probability
of 3\,$\times$\,$10^{-6}$. The other two noticable peaks at 370\un{s} and 850\un{s}
have respective probabilitites of 0.006 and 0.007, and are therefore not significant.
Peaks with probabilities greater than $10^{-3}$ cannot be considered significant
if we apply a 3$\sigma$ requirement to claim a detection and thus, no peaks 
in the periodograms other than the one at 1330\un{s} can be taken as significant.
The after-trial significance for this signal is $Y$\,=\,4.14, and using Eq.\,\ref{eq:deJager94} 
we find a pulsed fraction of 0.13 ($N$\,=\,2023): a very reasonable number,

Fig.\;\ref{f:phasogram} presents three phasograms resulting from folding the 
flare light curve with a phase of 1330\un{s}. These were constructed in the
2--4\un{keV}, 4--10\un{keV} and 2--10\un{keV} bands. There is a clear asymmetry 
in the overall light curve, and a marked peak present at low energies, making
the 2--4\un{keV} band radically different from the much smoother 
4--10\un{keV} light curve. This peak is shifted by $\sim$266\un{s} from the centre
of the sinusoid, corresponding to a distance of $\sim$0.5\un{AU} for light.
This intriguing feature will be investigated elsewhere.

Finally, we also analyzed the pre-flare period of the 2002 October 3 
observation to search for the periods reported by Aschenbach et al.\ (2004) 
in these data. The periodogram is shown in Fig.\;\ref{f:beforeFlare516}
and is remarkably flat over most of the range of physically interesting periods.
As we can see from the associated probabilities, 
there are no significant peaks present.

\section{Summary and Conclusion}
\label{s:summary}

We detected the presence of a 22.2\un{min} quasi-period in the \xray data of
the flaring event in \sgra that occured on 2004 August 31 while simultaneous
\xmm and HST observations were under way (see Yusef-Zadeh et al.\ 2006,
B\'elanger et al.\ 2005 and B\'elanger et al.\ 2006 for details on the 
results of the campaing). The probability that such a signal arises 
from statistical fluctuations was determined to be 3\,$\times$\,$10{-6}$ from MC
simulations and is conservatively estimated to be less than 
$\sim$7.2\,$\times$\,$10^{-5}$ if we consider the 24 IFS in the 
period range 300--1600\un{s} \citep{c:dejager89}.
It is important to clarify that although an evolution of the period as a function
of time may be present (as expected from the spiraling motion), 
and that it can be qualitatively seen in the light curve as alluded to by 
Liu et al.\ (2006), our statistics do not permit us to claim the detection of such a feature.

We considered and analysed two other \xmm observations of \sgra using the same
method for the probability estimation: the one that contains the  2004 March 31 flare, 
and the one of the 2002 October 3 flaring event. The former was found to be unreliable 
due to solar proton contamination, and no significant signals were detected in the latter,
in contrast with the results obtained by Aschenbach et al.\ (2004) using the same data set.

The method we have developed to detect a periodic signal, and 
evaluate the associated probability is well suited for searches involving long periods 
with respect to the total observation time, and in which fluctuations in the  average 
count rate do not follow Poisson statistics.  This MC based method is just as powerful in the 
absence of red noise but clearly not necessary since the probability can be computed 
analytically. This straightforward and reliable method for quantitatively estimating the 
significance of a given periodic or semi-periodic modulation will be useful in future 
applications of this kind, and hopefully help, at least in part, relieve some of the 
ambiguities that arise in this type of analysis.

The radius of the \mso in units of $r_g$\,=\,$GM/c^2$ is given by:
\begin{equation}
r_{\rm mso} = [3 + z_2 - \sqrt{(3 - z_1)\,(3 + z_1 + 2z_2)}],
\end{equation}
where $z_1 =  1 + (1-a^2)^{1/3} [(1+a)^{1/3} + (1-a)^{1/3}]$,
and $z_2 = (3a^2 + z_1^2)^{1/2}$.
The Keplerian frequency is:
\begin{equation}
\omega = \frac{c^3}{2\pi GM} (r^{3/2} + a)^{-1}.
\end{equation}
Taking a mass of 3.5\,$\times$\,$10^6$\msol and assuming that the
periodic signal originates from orbital motion at the \mso, the 1330\un{s} 
period implies that the BH is spinning at a rate given by $a$\,$\approx$\,0.22
(in the prograde case, the only considered).

Thus, the main result presented here can be interpreted as the signature of a transient
event that evolved over about 10\un{ks}, simultaneously giving rise to an \xray and
near-infrared flare. The emitting region spiraled around the mildly spinning 
SMBH, near or at the last stable orbit.
In this simple scenario, we would have directly witnessed the accretion of matter 
through the accretion disk in the Kerr metric around the central dark mass
(see Tagger \& Melia 2006).


\vspace{-5mm}
\acknowledgements{
	G.B.\ acknowledges the support of the French Space Agency}


\begin{thebibliography}{dummy}
\bibitem[Aschenbach et al.\ 2004] {c:aschenbach04} Aschenbach, B. et al.\ 2004, \aap, 417, 71
\bibitem[Baganoff et al.\ 2001] {c:baganoff01} Baganoff, F.K. et al.\ 2001, \nat, 413, 45 
\bibitem[Baganoff et al.\ 2003] {c:baganoff03} Baganoff, F.K. et al.\ 2003, \apj, 591, 891
\bibitem[B{\'e}langer et al.\ 2005] {c:belanger05} B{\'e}langer, G., Goldwurm, A. et al.\ 2005, \apj, 635, 1095
\bibitem[B{\'e}langer et al.\ 2006] {c:belanger06} B{\'e}langer, G., et al.\ 2006, \apj, 636, 275
\bibitem[Buccheri et al.\ 1983] {c:buccheri83} Buccheri, R., et al.\ 1983, \aap, 128, 245 
\bibitem[de Jager, Swanepoel \& Raubenheimer 1989] {c:dejager89} de Jager, O.C., Swanepoel, J.W.H. \& Raubenheimer, B.C.\ 1989, \aa, 221, 180
\bibitem[de Jager 1994] {c:dejager94} de Jager, O.C.\ 1994, \apj, 436, 239 
\bibitem[Genzel et al.\ 2003] {c:genzel03} Genzel, R. et al.\ 2003, \nat, 425, 934
\bibitem[Goldstein, Morris \& Yen 2004] {c:golstein04} Goldstein, M.L., Morris, S.A.\ \& Yen, G.G.\ 2004, (cond-mat/0402322)
\bibitem[Goldwurm et al.\ 2003] {c:goldwurm03} Goldwurm, A.\ et al.\ 2003a, \apj, 584, 751
\bibitem[Leahy et al.\ 1983] {c:leahy83} Leahy, D.A., Elsner, R.F. \& Weisskopf, M.C.\ 1983, \apj, 272, 256 
\bibitem[Liu, Melia \& Petrosian 2006] {c:liu06} Liu, S.\ Melia, F.\ \& Petrosian, V.\ 2006, \apj, 636, 789 
\bibitem[Porquet et al.\ 2003] {c:porquet03} Porquet, D. et al.\ 2003a, \aap, 407, L17
\bibitem[Tagger \& Melia 2006] {c:tagger-melia06} Tagger, M., \& Melia, F.\ 2006, \apjl, 636, L33
\bibitem[Vaughan 2005] {c:vaughan05} Vaughan, S.\ 2005, \aap, 431, 391
\bibitem[Yusef-Zadeh et al.\ 1999] {c:yusefZadeh99} Yusef-Zadeh, F., Choate, D., \& Cotton, W.\ 1999, \apjl, 518, L33
\bibitem[Yusef-Zadeh et al.\ 2006] {c:yusefZadeh06} Yusef-Zadeh, F.\ et al.\ 2006, \apj (in press)
\end{thebibliography}
\end{document}